\documentclass[]{tMOP2e}
\usepackage{amsfonts}

\newcommand{\dd}{\textrm{d}}
\newcommand{\re}{\textrm{Re}}
\newcommand{\im}{\textrm{Im}}
\newcommand{\lo}{\phi_{LO}}

\newcommand{\dz}[2]{\frac{\partial #1}{\partial #2}}

\newcommand{\tp}{\tau_P}

\newcommand{\avg}[1]{ \langle #1 \rangle}

\newcommand{\w}{\omega}

\newcommand{\f}{\varphi}

\newcommand{\ao}[2]{\hat{a}_{#1}(#2)}
\newcommand{\axo}[2]{\hat{a}^\dagger_{#1}(#2)}
  
\newcommand{\gc}{\langle  \hat{a}^\dagger \hat{a} \rangle}
\newcommand{\fc}{\langle  \hat{a} \hat{a} \rangle}

\newcommand{\axa}{ \langle \hat{a}^\dagger (t) \hat{a}(t')\rangle  }
\newcommand{\aag}{ \langle \hat{a}(t) \hat{a}(t') \rangle  }
\newcommand{\axaw}{\langle \hat{a}^\dagger (\omega) \hat{a}(\omega')\rangle  }
\newcommand{\aaw}{ \langle \hat{a}(\omega) \hat{a}(\omega') \rangle  }
\newcommand{\axap}[2]{ \langle \hat{a}^\dagger (#1) \hat{a}(#2)\rangle  }
\newcommand{\aap}[2]{ \langle \hat{a}(#1) \hat{a}(#2) \rangle  }

\newcommand{\xx}{ \langle \hat{x}^2 \rangle }
\newcommand{\xy}{ \langle \hat{x}^2_{\pi/2} \rangle }

\title{Multimode Spontaneous Parametric Down-Conversion\\ in the Lossy Medium}
\author{Rados\l aw Chrapkiewicz$^{\ast}$\thanks{$^\ast$Corresponding author. Email: radekch@fuw.edu.pl}, Wojciech Wasilewski
\\\vspace{6pt}  {\em{Institute of Experimental Physics, University of Warsaw, ul. Hoza 69, 00-681 Warsaw, Poland \\ phone: +48--22--55--32--120, fax: +48--22--625--64--06}}\\\vspace{6pt}\received{received: 21st October 2009}
}

\date{\small Dated: \today}
\numberwithin{equation}{section}
\begin{document}
\maketitle

\begin{abstract}
We study the process of multimode Spontaneous Parametric Down--Conversion (SPDC) in the lossy, one dimensional waveguide. We propose a description using first order Correlation Functions (CF) in the fluorescence fields, as a very fruitful and easy approach providing us with a complete information about the final multimode state. We formulate the equation of the evolution of the multimode CF along the crystal using four characteristic length scales. We solve it analytically in the one mode case and numerically in the multimode case. We capture simultaneous effects of three wave mixing with ultrashort pump, linear propagation and attenuation, and we are able to divide the evolution into three stages and predict it qualitatively. We find that losses do not destroy the quantum properties of SPDC but stabilize the final state.

\begin{keywords}
Multimode Spontaneous Parametric Down-Conversion; Squeezed states; Gaussian states of light; Dissipative dynamics; First order correlation functions.
\end{keywords}
\end{abstract}

\section{Introduction}
Optical Parametric Amplifiers (OPAs) are commonly used as a source of non-classic light. Operated below threshold they produce squeezed light which can be used in a vast variety of quantum protocols, including state teleportation \cite{Furusawa98}, quantum cryptography \cite{RalphPRA00}, interferometry \cite{VahlbruchPRL08} and precise measurements \cite{KuzmichPRL00}.

A typical OPA comprises pumped nonlinear crystal placed in an optical cavity resonant at exactly half the pump frequency. In such systems narrowband, continuous-wave squeezing is obtained \cite{SchoriPRA02}. In this paper we focus on the other possibility, when the nonlinear medium is pumped with a short pulse and a broadband squeezed light is produced in a single pass through the crystal in a process commonly called Spontaneous Parametric Down-Conversion (SPDC). This is especially efficient when a waveguide made in the nonlinear crystal is used to carry the pump and squeezed light \cite{AndersonOL95}. 

It is well known that optical losses destroy squeezing. However nonclassical character of those states still manifests itself in residual squeezing present even after severe attenuation. This behavior is rather uncommon among nonclassical states of light. Most of them die rapidly subject to deleterious contact with environment. Therefore we have chosen to study the effects of distributed losses, unavoidable in realistic nonlinear materials, on generation of squeezed light. It turns out that squeezing can still be generated in lossy media and it is even stabilized under certain conditions. However, the output state no longer saturates the uncertainty principle.

In previous works the multimode state of squeezed light was described either with help of Green functions \cite{WasilewskiPRA06A} or perturbative expansions \cite{Keller97} have been used. However those approaches cannot be used for numerical description of the lossy media, because additional, very large Green functions appear. Therefore we have adopted a formalism which traces the evolution of first order correlation functions of the fluorescence fields. This approach is computationally effective and provides complete information about the final multimode state. Moreover the evolution of the correlation functions is simple and qualitatively easily predictable. Below we will describe various possible cases and assess characteristic widths of the correlation functions by pump pulse length, crystal thickness and its fundamental dispersive properties.

An SPDC process in an ideally transparent nonlinear medium results in creation of multiple independent squeezed states \cite{BraunsteinBM,WasilewskiPRA06A}. They occupy orthogonal modes of light field and can be in principle separated \cite{OpatrnyPRA2002}. Moreover each of the states is pure and saturates Heisenberg inequality for variance of $\hat x$ and $\hat p$ quadratures. 

However, as soon as loses come into play full description of the final state becomes much more complicated. The state can be cast into Williamson form \cite{CarusoNJP08}, which corresponds to a reconstruction  as a multimode squeezed state subject to losses mixing components with various squeezing. Unfortunately  there is no way to separate the field into uncorrelated modes. One can find a set of polychromatic quadratures of light $\hat{x}_i$ that are not correlated with any other commuting quadratures, however together with they conjugates they exceed minimal uncertainty bound. 

We describe the multimode state of light in terms of its experimentally important characteristics. We trace the evolution of the average number of photons, maximum squeezing and the product of variances of maximally squeezed quadrature and its conjugate.
We differentiate three general stages of the evolution: an exponential growth and a squeezing, a linear temporal expansion without squeezing and a stabilization. We show that the boundaries between those stages typically correspond to characteristic lengths of either attenuation or pump-SPDC overlap.

The paper is organized as follows. In Sec. 2. we introduce the correlation function and describe the adopted model of OPA. In Sec. 3. we present results of numerical simulations of the evolution of the correlation functions. In Sec. 4. we connect the correlation functions with the results of the Homodyne Detection (HD). By performing an optimization procedure we find the most squeezed mode of the light. Furthermore, we gather the results of our simulations and divide the evolution into stages. Finally  Sec. 5. concludes the paper.

\section{Model}

In this section we develop a model of a lossy OPA. In particular we introduce correlation functions and equations for their evolution. We define four characteristic length scales characterizing the evolution of the SPDC light.

We will focus on interactions in a waveguide made in a nonlinear medium \cite{AndersonOL95} in which fields propagate along $z$ axis in fixed spatial modes. We assume that the interaction of the pump at frequency $2\omega_0$ and the fluorescence at frequency $\omega_0$ is perfectly phase matched.  Due to the small intensity of light in SPDC the pump remains undepleted.
We will first discuss a situation in which only the SPDC undergoes linear losses. Then we show how the evolution is modified when the pump is also damped. 

We begin by discussing the propagation of the field at a degereate frequency $\omega_0$ in a lossless OPA. The equation of evolution of the SPDC annihilation operator $\hat a$ can be written as $\partial \hat{a} / \partial \textrm{z}=\hat{a}^\dagger/L_{NL}$, \cite{BoydNLO} where characteristic length of the nonlinear coupling $L_{NL}$ is inversely proportional to the pump amplitude $P_0$ and nonlinear coupling. Let us note, that this equation is formally identical to the equation of evolution of the classical field amplitude along the nonlinear medium.

\begin{figure*}
 \centering
 \includegraphics[width=0.8 \textwidth]{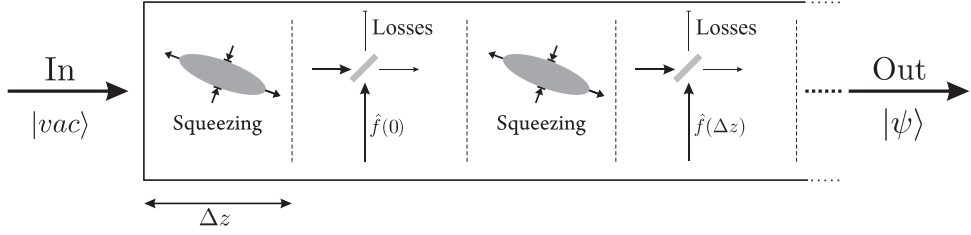}
 \caption{Schematic representation of the model of lossy OPA. The OPA is divided into thin slices of thickness $\Delta z$}
 \label{fig:slice}
\end{figure*}

For a lossy nonlinear medium the description becomes more involved. Classical equation of motion would in such case contain a term $-\hat a/L_A$ with $L_A$ being attenuation length. In quantum case this term is inevitably accompanied by an additional vacuum noise:
\begin{equation}
\dz{\hat{a}}{z}=-\frac{\hat{a}}{L_A} + \frac{\hat{a}^\dagger}{L_{NL}}+ \sqrt{\frac2{L_A}} \hat{f}(z)
\label{aevol}
\end{equation}
where $\hat{f}(z)$ are Langevin noise operators. They form a set of independent bosonic operators indexed by $z$, i.e.: $[\hat{f}(z),\hat{f}^\dagger(z')]=\delta(z-z')$ representing vacuum modes $\avg{\hat f(z)}=0$, $\avg{\hat f^\dagger(z)\hat f(z)}=\avg{\hat f(z)\hat f(z)}=0$. The above evolution can be understood as an alternate application of squeezing and beamsplitter type losses, as illustrated in Fig.~\ref{fig:slice}.
Eq. \eqref{aevol} can be integrated to the following input-output relations:
\begin{multline}\label{agreen}
\hat{a}(L)=e^{-L/L_{A}}\left(\hat{a}(0)\cosh L/L_{NL}+\hat a^\dagger(0)\sinh L/L_{NL}\right)  \\
+ \sqrt{\frac{2}{L_A}} \int_0^L \textrm{d}r \ e^{-r/L_{A}}\left(\hat f(L-r)\cosh r/L_{NL}+\hat f^\dagger(L-r)\sinh r/L_{NL}\right)
\end{multline}
Note that the solution involves entire set of $\hat{f}(z)$, although one expects that the final state on the output could be described without tracing the nuances of this dependence.  Indeed, we may use the fact, that the state of SPDC is always
Gaussian centered around the origin of the phase space.
This is because we start from vacuum states that also have this
property and the equations of motion \eqref{aevol} are linear, corresponding
to linear distortion of the phase space. Therefore the final state is completely described  by its second order moments $\avg{\hat{a}^\dagger \hat{a}}$ and $\avg{\hat{a} \hat{a}}$. Both of them are $c$-numbers and their values can be easily traced by solving their evolution equations. They are in turn found by inserting Eq.~\eqref{aevol} into the $z$-derivatives of the second order moments:
\begin{subequations}
  \begin{align}
    \dz{\gc}{z}&=\frac{2}{L_{NL}} \re \fc  -\frac{ \gc}{L_A} 
 \label{eq:ev1G}\\
 \dz{\fc}{z}&=\frac{2}{L_{NL}}\gc-\frac{ \fc}{L_A} +\frac{1}{L_{NL}}  \label{eq:ev1F}
  \end{align}
\end{subequations}
Note, that in the above equations the Langevin noise operators $\hat f(z)$ do not appear, because their correlation functions are zero. The above equations can be solved analytically, completely describing evolution of the state of SPDC along the amplifier:
\begin{subequations}
  \begin{align}
\axap{L}{L}&=\frac{1}{L_{NL}}\frac{1}{L_A^{-2}-4 L_{NL}^{-2}}\left(\frac{2}{L_{NL}}
-e^{-L/L_A}\Big(\frac{1}{L_A} \sinh \frac{2L}{L_{NL}} + \frac{2}{L_{NL}}\cosh \frac{2L}{L_{NL}}\Big)\right) \label{eq:analsol1} \\
\aap{L}{L}&=\frac{1}{L_{NL}}\frac{1}{L_A^{-2}-4L_{NL}^{-2}}\left(\frac{1}{L_A}
-e^{-L/L_A}\Big(\frac{2}{L_{NL}} \sinh \frac{2L}{L_{NL}} + \frac{1}{L_A} \cosh \frac{2L}{L_{NL}}\Big)\right) \label{eq:analsol2}
  \end{align}
 \label{eq:analsol}
\end{subequations}
In principle the same result can be obtained from the Heisenberg transformation \eqref{agreen}, however this way we would  first laboriously find the entire history of noise contribution only to trace it away in the next step. 

Let us briefly discuss this result, as it also captures important features of a full multimode evolution. 
For $L_A>L_{NL}/2$ both $\avg{\hat{a}^\dagger \hat{a}}$ and $\avg{\hat{a} \hat{a}}$ grow exponentially. 
However, for losses exceeding critical level $L_A<L_{NL}/2$, sinh and cosh functions are dwarfed by the attenuation exponent and the $\avg{\hat{a}^\dagger \hat{a}}$ and $\avg{\hat{a} \hat{a}}$ ultimately reach steady state values.

When the pump is a short pulse  and we wish to include material properties of the medium, a multimode description of the SPDC by a family of operators $\hat a(\omega)$ becomes necessary. The dispersion of the crystal is reflected in the dependence of the wavenumber on frequency $k(\omega)$. It is included in the propagation equations by adding a term of type $ik(\omega)\hat a(\omega)$. Since the spectrum of SPDC in a waveguide is restricted by  moderate phasematching bandwidth, we can expand $k(\omega)$ around SPDC central frequency $\omega_0$ up to second order. The expansion coefficients are inverse phase velocity, inverse group velocity and second order dispersion respectively. Similar expansion around the pump frequency $2\omega_0$ yields its propagation parameters. We assume that the pump bandwidth is sufficiently narrow so that the effects of the second order dispersion on the pump pulse lengthening can be neglected. Due to phase matching assumption phase velocities of the SPDC and pump match and drop out. Therefore, to describe the material properties,  it is sufficient that we know difference of inverse group velocities between the pump and the SPDC $\Delta \beta_1$ and dispersion coefficient for SPDC $\beta_2$:
\begin{align}
\Delta \beta_1&=\left.\frac{\textrm{d}k}{\textrm{d}\omega}\right|_{\omega=\omega_0}-
\left.\frac{\textrm{d}k}{\textrm{d}\omega}\right|_{\omega=2\omega_0}, &
\beta_2&=\left.\frac{\textrm{d}^2k}{\textrm{d}\omega^2}\right|_{\omega=\omega_0}.
\end{align}

We assume the spectral amplitude $P(\omega)$ of the pump pulse to be Gaussian corresponding to a duration $\tau_P$ and a maximum amplitude $P_0$
 \begin{equation}
  P(\omega)=P_0 \ e^{-(\omega-2\omega_0)^2 \tau_P^2/2}.
 \end{equation}
Now we can introduce a length scale $L_{OV}$ over which pump pulse and the SPDC light overlap $L_{OV}=\tau_P/\Delta\beta_1$.
Additionally one can define a crystal length $L_D$ over which the spectrum of the SPDC would get as narrowband as the pump spectrum, $L_D=\tau_P^2/\beta_2$ \footnote{See Sec. 3 for detailed explanation of the role of $L_D$}. With those definitions at hand we can write a multimode equation of the propagation of $\hat{a}(\w)$ in the reference frame of the pump pulse:
\begin{equation}
\dz{\hat{a}(\omega)}{z}=  i\left(\frac{\tp \omega}{L_{OV}}+\frac{\tp^2 \omega^2}{2 L_D}\right)\hat{a}(\omega)+\frac{1}{P_0 L_{NL}} \int P(\omega+\omega') \hat{a}^\dagger (\omega') \ \dd \omega'
-\frac{1}{L_A} \hat a(\omega) + \sqrt\frac{2}{L_A} \hat f(\omega)
 \label{eq:a_evol}
\end{equation}
The first term represents linear dispersive evolution in the pump reference frame. The second term describes nonlinear interaction, which couples SPDC components of frequency $\omega$ and $\omega'$ with pump components of frequency $\omega+\omega'$. The last two terms describe losses.

Since the above equation is linear its solution can be always expressed as an integral transform:
\begin{multline}
 \hat{a}(\omega,z)=\int C(\omega,\omega') \hat{a}(\omega',0) \ \textrm{d}\omega'
+\int S(\omega,\omega') \hat{a}^\dagger(\omega',0) \ \textrm{d}\omega' \\
+\int \textrm{d}\omega' \textrm{d}z' C_f(\omega,\omega',z') \hat f(\omega',z') 
+\int \textrm{d}\omega' \textrm{d}z' S_f(\omega,\omega',z') \hat f^\dagger(\omega',z') 
\end{multline}
The Green functions $C(\omega,\omega'),S(\omega,\omega'),C_f(\omega,\omega',z'),S_f(\omega,\omega',z')$ contain all the information about the amplifier, in particular they are sufficient to describe the final state. However, finding numerical approximations of $C_f(\omega,\omega',z')$ and $S_f(\omega,\omega',z')$ would be very tedious and unnecessary. The quantum state of SPDC is fully described by a  multimode generalizations of the second order moments which become Correlation Functions (CF) $\axaw$, $\aaw$. They express the amplitude and the phase correlation of the light between two arbitrary frequencies $\omega$ and $\omega'$. 
The evolution of the CF can be traced by solving their evolution equations. These are found by differentiating $\axaw$ and $\aaw$ over $z$ and using Eq.~\eqref{eq:a_evol}
\begin{subequations}\label{eq:evGF}
   \begin{align}
   \dz{\axap{\w}{\w'}}{z}&=\left(L^{*}(\w)+L(\w')\right)\axap{\w}{\w'}-\frac{1}{L_A} \axap{\w}{\w'} \nonumber \\
&+ \frac{1}{P_0 L_{NL}}\int \dd \omega'' \left [ P^*(\w'')\aap{\w-\w''}{\w'}+P(\w'')\aap{\w}{\w''-\w'}^* \right] \label{eq:evGs}\\
\dz{\aap{\w}{\w'}}{z}&=\left(L(\w)+L(\w')\right)\aap{\w}{\w'}-\frac{1}{L_A} \aap{\w}{\w'} + \frac{1}{P_0 L_{NL}} P(\w'+\w)  \nonumber\\
&+\frac{1}{P_0 L_{NL}}\int \dd \omega'' \left [P(\w'')\axap{\w''-\w}{\w'}+P(\w'')\axap{\w}{\w''-\w'}^* \right] \label{eq:evFs}.
  \end{align}
\end{subequations}
Above $L(\omega)$ denotes the linear propagation operator in frequency domain
$  L(\w)= i \tp \w/L_{OV}  + i \tp^2 \w^2 / (2 L_D)  $.

Actually it is more intuitive to trace  the evolution of the CF in the time domain, rather than in frequencies. It is meaningful to apply double Fourier transform to $\axaw$ and $\aaw$. This way we obtain CF in time:
\begin{align}
\axa &=\int \textrm{d}\w \textrm{d}\w ' e^{i\w t-i\w ' t'}\axaw, &
\aag &=\int \textrm{d}\w \textrm{d}\w ' e^{-i\w t-i\w ' t'}\aaw.
\end{align}
This transform can be applied to both sides of \eqref{eq:evGF} and one can obtain evolution equations for CF in time domain: 
\begin{subequations}\label{eq:evtGF}
   \begin{align}
   \dz{\axa}{z}&=\left(\mathcal L^{*}_t+\mathcal L_{t'}\right)\axa-\frac{\axa}{L_A} 
+ \frac{P^*(t)\aag+\aag^* P(t')}{P_0 L_{NL}}\label{eq:evtGs}\\
\dz{\aag}{z}&=\left(\mathcal L_t+\mathcal L_{t'}\right)\aag-\frac{\aag}{L_A}  \nonumber\\
&+\frac{P(t)\axa+\axa^* P(t')}{P_0 L_{NL}}+ \frac{P(t) \delta(t-t')}{P_0 L_{NL}}  
\label{eq:evtFs}
  \end{align}
\end{subequations}
The first term in \eqref{eq:evtGF} represents time shift and dispersive broadening, the second term represents attenuation and the last term represents nonlinear interaction.
The operators $\mathcal L_t$ represents convolution with a Fourier transform of $L(\w)$ and can be rewritten as:
\begin{align}
\mathcal{L}^*_t+\mathcal{L}_{t'}&=
\frac{2\tp}{L_{OV}}\frac{\partial}{\partial(t+t')}+\frac{2i\tp^2}{L_{D}}\frac{\partial^2}{\partial(t+t')\partial(t-t')} \nonumber\\
\mathcal{L}_t+\mathcal{L}_{t'}&=
\frac{2\tp}{L_{OV}}\frac{\partial}{\partial(t+t')}
-\frac{i\tp^2}{L_{D}}\left(\frac{\partial^2}{\partial(t+t')^2}+\frac{\partial^2}{\partial(t-t')^2} \right)
\end{align}
where we on purpose changed the variables, so that differentiation along the diagonal and antidiagonal is emphasized.

\section{Evolution of the correlation functions}

Correlation Functions (CF)  $\axaw$ and $\aaw$ completely describe the Gaussian quantum state of the SPDC. 
Therefore it is beneficial to understand evolution of the CF along the crystal. Despite complicated form of the Eqs.~\eqref{eq:evGF}, we will show that this evolution can be intuitively understood  and the CF shape is directly connected with four characteristic length scales: pump-SPDC overlap length $L_{OV}$, dispersive length $L_D$, attenuation length $L_A$ and nonlinear length $L_{NL}$. $L_{OV}$ sets the effective length of the crystal over which the nonlinearity can build up SPDC in an undisturbed way. $L_D$ is the characteristic length over which correlations between different moments in time arise. Finally $L_A$ sets the crystal length over which contributions from the front slices of the crystal are erased and the state stabilizes.

We have calculated the CF in the subsequent distances of the propagation by numerically solving matrix approximations of Eqs. (\ref{eq:evGF})  using the Split Step Fourier Method. We have verified that our simulation is accurate by repeating it on grids finer in each dimension. Calculations were carried out for a comprehensive set of the characteristic lengths in order to understand a mechanisms governing the evolution of CF. Our findings are most conveniently described in the time domain.

 \begin{figure}
\subfigure[$|\axa|$ at $L=L_{NL}/20$]{\label{fig:cf:a}\includegraphics[width=7cm]{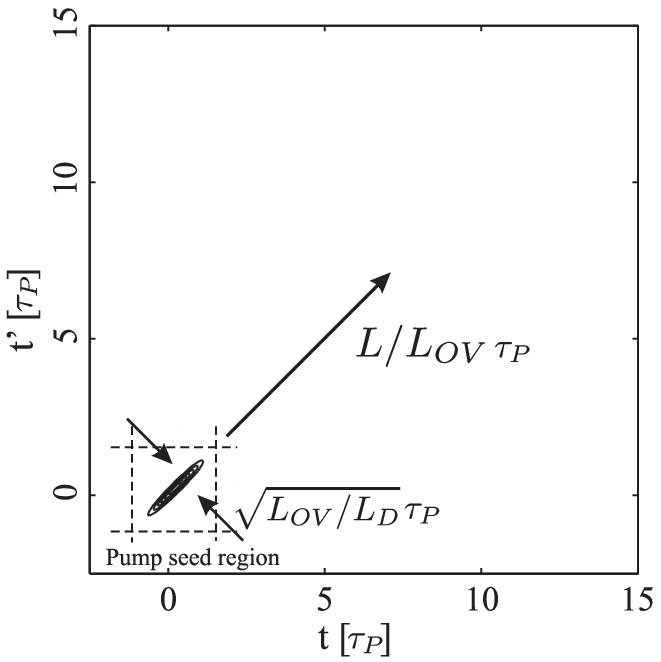}}
\subfigure[$|\aag|$ at $L=L_{NL}/20$]{\label{fig:cf:b}\includegraphics[width=7cm]{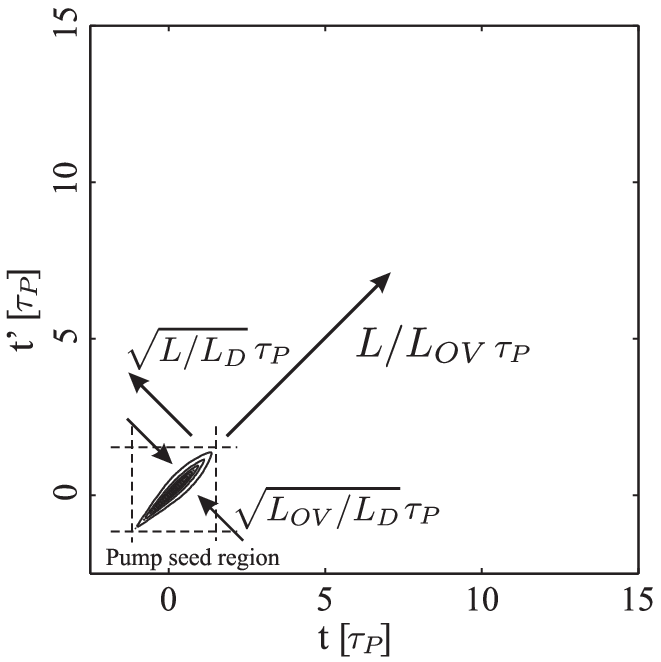}}\\
\subfigure[$|\axa|$ at $L=L_{NL}/3$]{\label{fig:cf:c}\includegraphics[width=7cm]{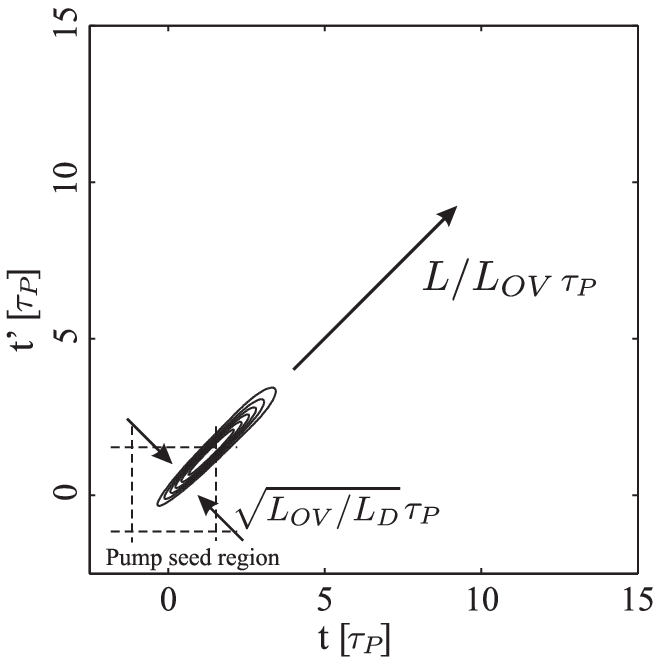}}
\subfigure[$|\aag|$ at $L=L_{NL}/3$]{\label{fig:cf:d}\includegraphics[width=7cm]{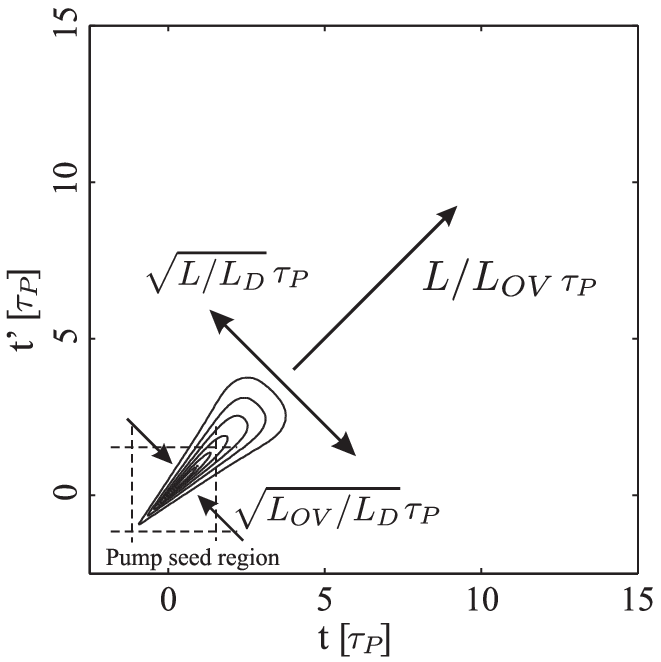}}\\   
\subfigure[$|\axa|$ at $L=L_{NL}$]{\label{fig:cf:e}\includegraphics[width=7cm]{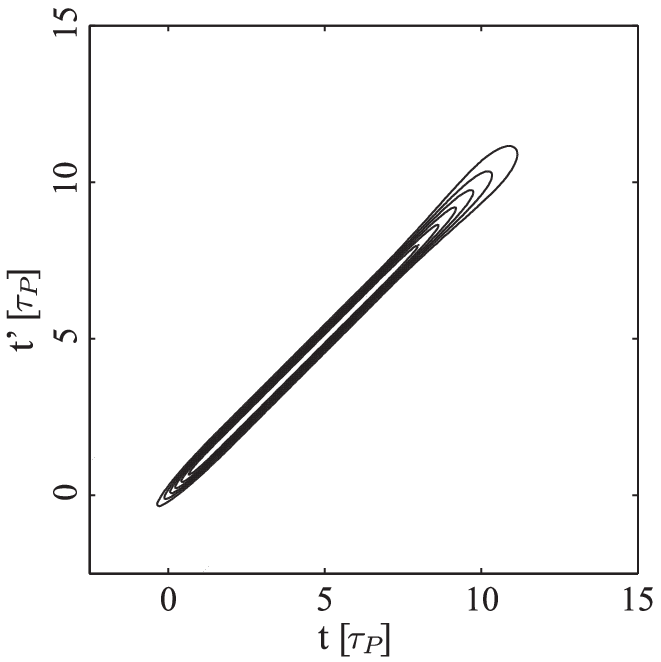}}
\subfigure[$|\aag|$ at $L=L_{NL}$]{\label{fig:cf:f}\includegraphics[width=7cm]{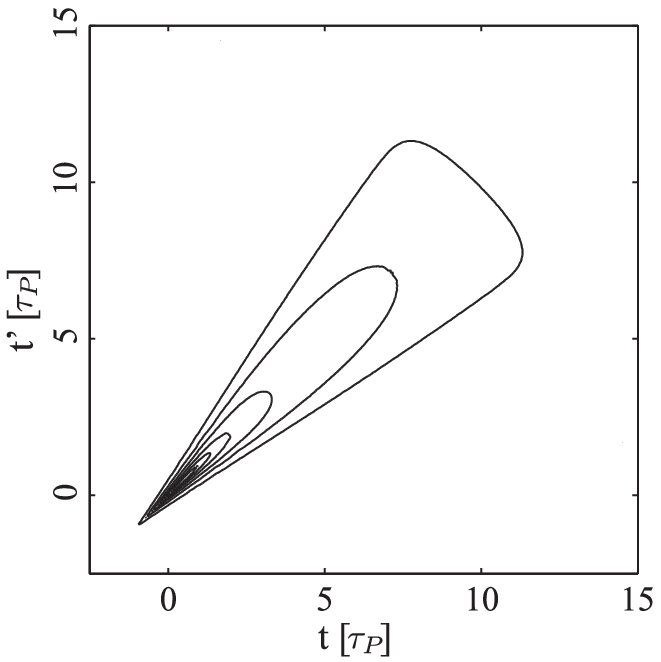}}\\
\caption{The contour plots of two Correlation Functions (CF) moduluses: $|\axa|$ and $|\aag|$ presented on  the $t - t'$ plain. We draw a lossless case for the pump-SPDC overlap length $L_{OV}= L_{NL}/10$ and dispersive length $L_D=3 L_{NL}$. The following pairs of plots present the evolution of the CF on the subsequent distances:  $L=L_{NL}/20$, $L_{NL}/3$ and $L_{NL}$.
}
{\label{fig:cf}}
\end{figure}

\begin{figure}
\subfigure[$\textrm{Arg}(\axa)$ at $L=L_{NL}$]{\includegraphics[width=7cm]{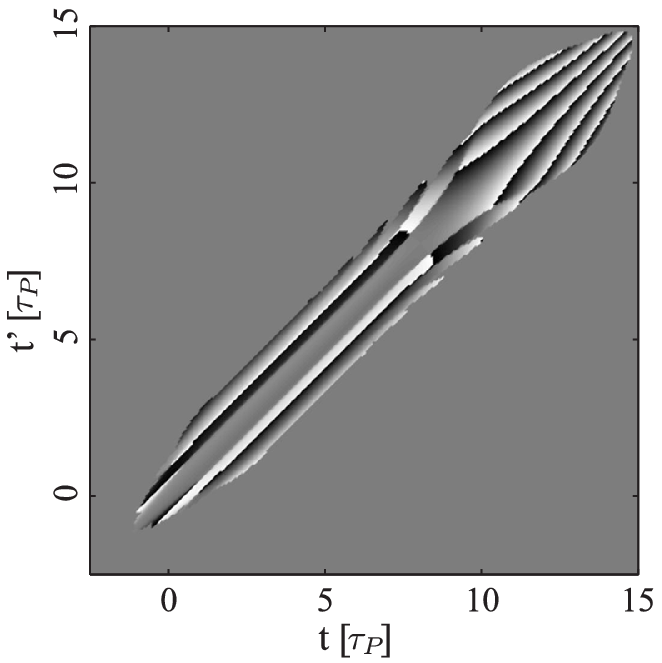}}
\subfigure[$\textrm{Arg}(\aag)$ at $L=L_{NL}$]{\includegraphics[width=7cm]{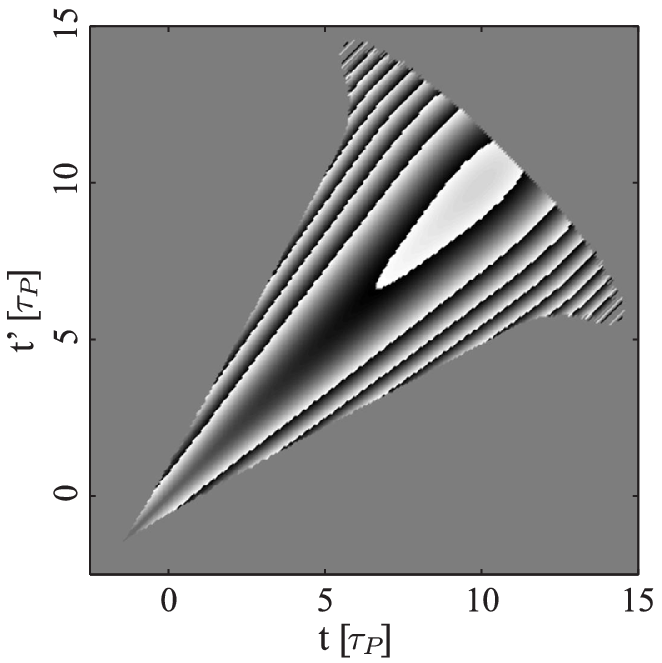}}
\caption{The plots of two Correlation Functions (CF) phase: $\textrm{Arg}(\axa)$ and $\textrm{Arg}(\aag)$ presented on  the $t - t'$ plain. We draw these plots for the same parameters as in the  Fig. \ref{fig:cf:e} and \ref{fig:cf:f}: $L_{OV}= L_{NL}/10$, $L_D=3 L_{NL}$,  $L=L_{NL}$.}
\label{fig:cf:ph}
\end{figure}

Firstly, let us discuss the lossless case, i.e. $L_A \rightarrow \infty$. In the Fig.~\ref{fig:cf} we present results of the numerical simulations performed with the following parameters:  pump-SPDC overlap length $L_{OV}= L_{NL}/10$ and dispersive length $L_D=3 L_{NL}$. Such choice gives an illustrative insight into a typical evolution of the CF inside the crystal when both a linear propagation and nonlinear coupling play comparable role.  Using contour plots in the Fig. \ref{fig:cf} we drew snapshots of the absolute value of the CF in the time domain $|\axa|$ and $|\aag|$. Snapshots were taken at crystal lengths $L=L_{NL}/20$, $L_{NL}/3$ and $L_{NL}$. In the Fig. \ref{fig:cf:ph} we draw contours of the phase of the CF at $L=L_{NL}$.
 
Let us analyze the plots presented in the Fig. \ref{fig:cf}. 
In the absence of losses the evolution is determined by the interplay between two terms of Eqs.~\eqref{eq:evtGF}: linear, responsible for the shifting and spreading of the CF and nonlinear, responsible for the growth of the CF. On a two-dimensional plane of $t$ and $t'$ the nonlinear terms act only in the region where $P(t)$ and $P(t')$ are nonzero. In particular the only seed term $P(t)\delta(t-t')$ is located on the diagonal. Over distances when dispersion is negligible $L\ll L_{OV}, L_D$, the correlation functions grow on the diagonal according to single mode solutions \eqref{eq:analsol}. Then they diverge due to material dispersion. Hence, after $L=L_{NL}/20$ of the propagation $\aag$ arises around the diagonal pump seed region (the diagonal of the square in the plots) as seen in the Fig.~\ref{fig:cf:b}. At the same time the other correlation function $\axa$ is seeded by the term involving the pump and $\aag$. 

Let us note a finite antidiagonal width of both $\axa$ and $\aag$. It is an effect of the second order dispersion of the crystal i.e. finite $L_D$. Indeed along the antidiagonal alone $t+t'= \text{const}$ the linear part of the equation of evolution of $\aag$ \eqref{eq:evtFs} has a form of a diffusion equation $\partial\aag/\partial z \propto \partial^2\aag/\partial (t-t')^2$
with a diffusion coefficient inversely proportional to $L_D$. Therefore the $\aag$ spreads in the antidiagonal direction and its width can be estimated as $\tau_p\sqrt{L/L_D}$. Within the pumped region where $P(t)$ and $P(t')$ are nonzero, this spread is then imprinted onto $\axa$ due to nonlinear term present in Eq.~\eqref{eq:evtGs}. 
Note that the antidiagonal width of the $\axa$ is inversely proportional to the diagonal width of $\axaw$. The latter is also the spectral bandwidth of SPDC which is now estimated as $\tp^{-1}\sqrt{L_D/L}$ for short crystal lengths.

In the next two plots, Fig. \ref{fig:cf:c} and \ref{fig:cf:d} we capture CF after an evolution distance $L_{NL}/3$. The influence of the group velocity difference between the pump and SPDC becomes apparent. In Eqs.~\eqref{eq:evtGF} it causes a shift of $\axa$ and $\aag$ along the diagonal with a speed of $\tau_p/L_{OV}$. Both CF unwind from the pump seed region. Let us note that $\aag$ is under continuous influence of dispersion caused diffusion in the antidiagonal direction. It assumes triangular shape with broad part corresponding to photon pairs born in the front slices of the crystal and narrow part corresponding to the photons produced at the end of the crystal. On the other hand the antidiagonal width of $\axa$ is limited to maximum width of $\aag$ in the pump seed region, which is estimated as $\tau_p\sqrt{L_{OV}/L_D}$. Therefore as soon as the crystal length exceeds $L_{OV}$ the spectrum of the SPDC stops shrinking. 

The Figures \ref{fig:cf:e} and \ref{fig:cf:f} show the CF in the later moments of evolution $L=L_{NL}$. All the above mentioned processes continue to shape the CF. The Fig.~\ref{fig:cf:ph} presents contour plots of phases of $\axa$ and $\aag$ in the region of significant absolute values. They describe temporal chirp acquired by photon pairs in the propagation through the dispersive crystal. 

  In the next step losses were added to the simulation. As we argued above for short propagation distances $L\ll L_{OV}$ one can apply one mode solution given in Eq.~\eqref{eq:analsol}. For longer distances two situations are possible. For high attenuation $L_{NL}>2 L_A$ and both of them much shorter than the overlap length $L_{OV}$ the CF will reach steady state values due to attenuation alone and dispersive effects will play a minor role. Therefore we show the results of the calculations only in the opposite regime, for $L_A=1/5 L_{NL}>L_{OV}$. In this case the main role of losses is attenuation of the tail of CF which continuously unwinds from pump seed region. Thus the diagonal length of the CF is restricted to the length of the order of $\tp L_A/L_{OV}$ as plotted in Fig. \ref{fig:cf}.
\begin{figure}
\subfigure[$\axa$]{\includegraphics[width=7cm]{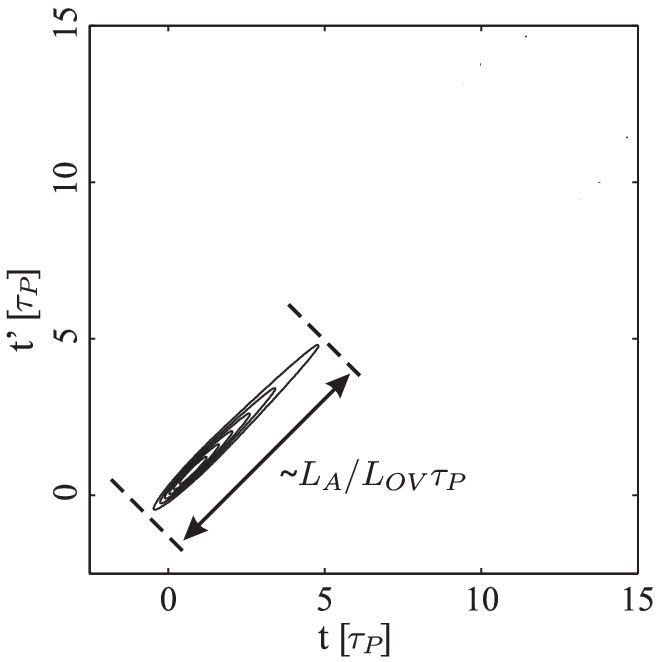}}
\subfigure[$\aag$]{\includegraphics[width=7cm]{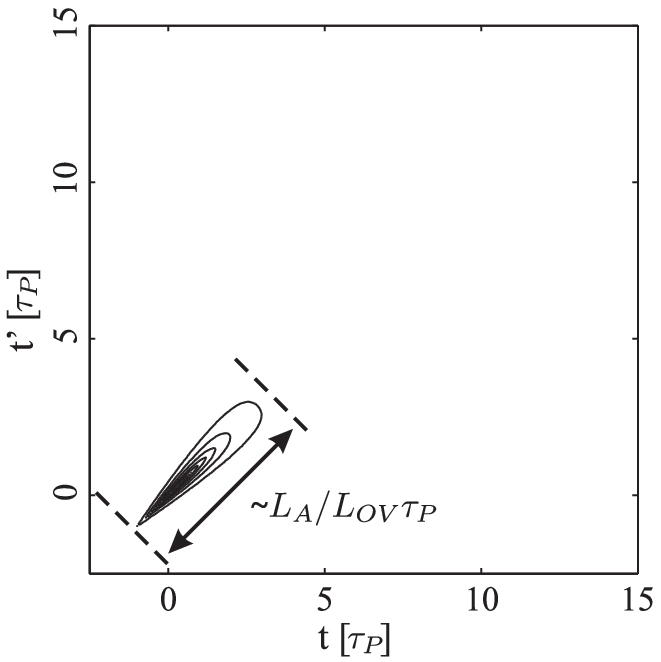}}
\caption{
The contour plots of two Correlation Functions (CF) moduluses: $|\axa|$ and $|\aag|$ presented on  the $t - t'$ plain. We draw the lossy case (the attenuation length $L_A=1/5 L_{NL}$) corresponding to the other parametes as in the   Fig. \ref{fig:cf:e} and \ref{fig:cf:f}:  $L_{OV}= L_{NL}/10$, $L_D=3 L_{NL}$,  $L=L_{NL}$. CF achieved a stationary state being unaffected during further propagation.}
\label{fig:cf:loss}
\end{figure}
The evolution at the distances $L\ll L_A$ is very similar to lossless case. Then the CF stabilize until finally for $L\gg L_A$ they reach steady state when their tail is fully stretched and cut by attenuation.

\section{Evolution of the quadrature variances}

OPA produces light which exhibit squeezing of quadrature variances below the shot noise level. Such states demonstrate non-classical properties. Applying the Homodyne Detection (HD) to the light generated in SPDC we can directly observe the amount of noise in the mode of a local oscillator. Similarly to the previous sections, it will be instructive to start with the analysis of the single mode OPA. In a single mode HD we measure the quadrature  $\hat{x}=\frac{1}{\sqrt{2}}\left( \hat{a} e^{i\f}+\hat{a}^\dagger e^{-i\f} \right)$, where $\f$ denotes the phase difference between the pump and the local oscillator. In the SPDC light the mean value of  $\hat{x}$ is zero. The variance of $\hat{x}$ which measures the quadrature noise equals
\begin{equation}
\xx=\frac{1}{2}+\gc+\re\{\fc e^{2i\f}\}
\label{quad1mode}
\end{equation}
We are interested in the most squeezed and most antisqueezed quadratures. In the lossless case we can use \eqref{eq:analsol} and \eqref{quad1mode} to retrieve a well-known result for the minimal and maximal variances in a single mode squeezing $\xx=e^{\pm 2 L/L_{NL}}/2$. The inclusion of losses characterized by attenuation length $L_A$ changes this expression into:
\begin{equation}
\xx=\frac{1}{2}+\frac{L_A}{2L_A\pm L_{NL}}\left(e^{-(L_A^{-1}\pm 2L_{NL}^{-1}) L}-1\right)
\label{varlosses}
 \end{equation}
We see that depending on the relation between $L_A$ and the nonlinear length $L_{NL}$ we deal with two different types of evolution. In the high losses regime $L_A<L_{NL}/2$ both minimal and maximal variances reach steady state. On the other hand in case of small losses $L_A>L_{NL}/2$ the antisqueezed quadrature grows exponentially, while the minimal $\xx$ decreases to a steady state value --- the squeezing is limited.
The above formula holds also in multimode case for crystal lengths much shorter than overlap length $L_{OV}$.

Let us follow with a multimode generalization of the previous discussion. We write down the expression for the quadrature $\hat{x}$ in the mode $\lo(\omega)$ given by the macroscopic field of Local Oscillator (LO).
\begin{equation}
  \hat{x}=\frac{1}{\sqrt{2}}\int \dd \omega \ \left[ \lo^*(\omega) \ao{}{\omega}+\lo(\omega) \axo{}{\omega} \right]
\end{equation} 
Variance of this quadrature $\xx$ can be obtained in a direct calculation:
 \begin{eqnarray}
  \left \langle \hat{x}^2 \right \rangle=
  \frac{1}{2} +\iint  \dd \omega \dd \omega' \left[ \lo(\omega) \axaw \lo^*(\omega') \right. \nonumber \\
   + \re\left(\lo(\omega) \aaw \lo(\omega') \right) \Big]
  \label{hom1}
\end{eqnarray} 
$\left \langle \hat{x}^2 \right \rangle=\frac{1}{2}$ refers to the shot noise level. Let us note that first  term under the integral does not depend on the phase of the $\lo$, while the second term containing $\aaw$ is responsible for phase sensitivity of the noise. 

In most applications of SPDC we desire to select the most squeezed mode. Such a mode, determined by appropriate LO function $\lo$, minimizes \eqref{hom1} for given CF $\axaw$ and $\aaw$. Since we apply the matrix approximation of the correlation functions, the integral in (\ref{hom1}) can be written as a quadratic form, multiplied both side by vectors representing real and imaginary part of $\lo$. The most squeezed mode is an eigenvector of the quadratic form corresponding to the smallest eigenvalue. This is described in details in the Appendix.

\begin{figure}
\subfigure[~]{ \label{fig:fmod} \includegraphics[width=8cm]{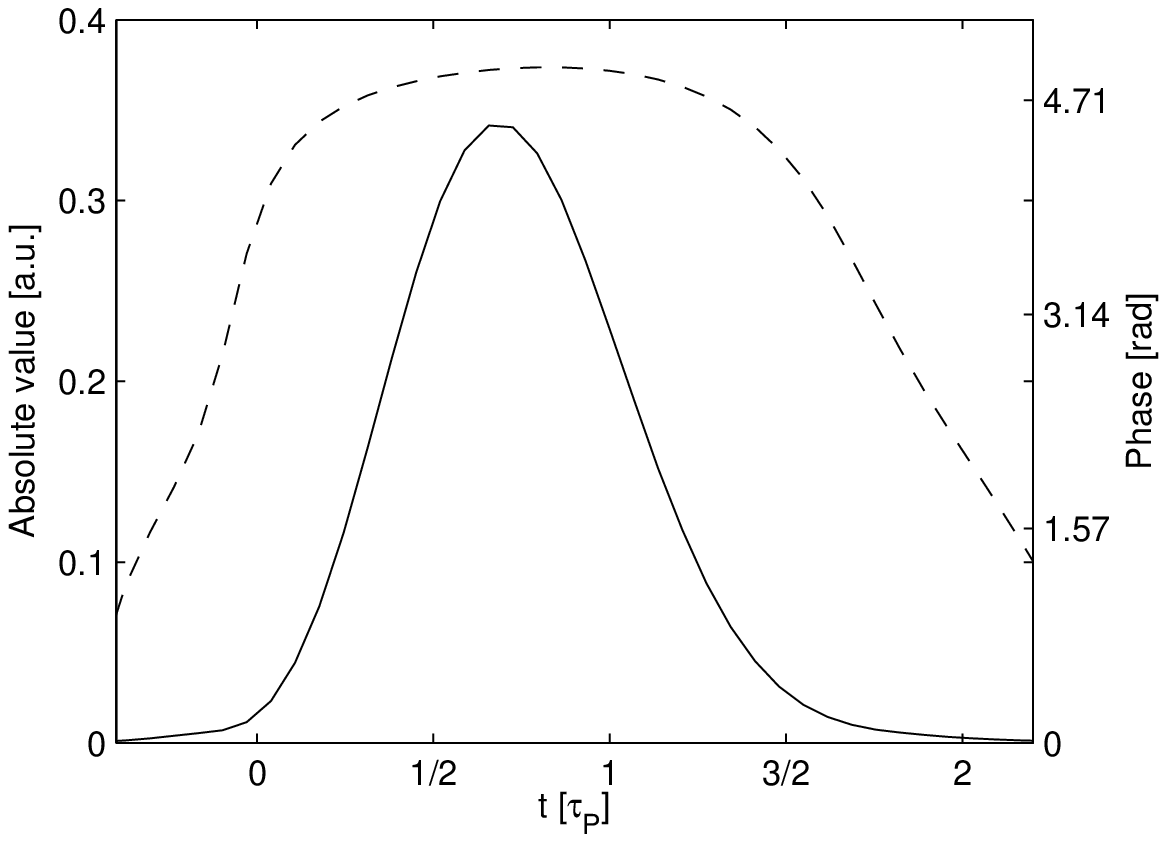}}
\subfigure[~]{ \label{fig:mbar} \includegraphics[width=8cm]{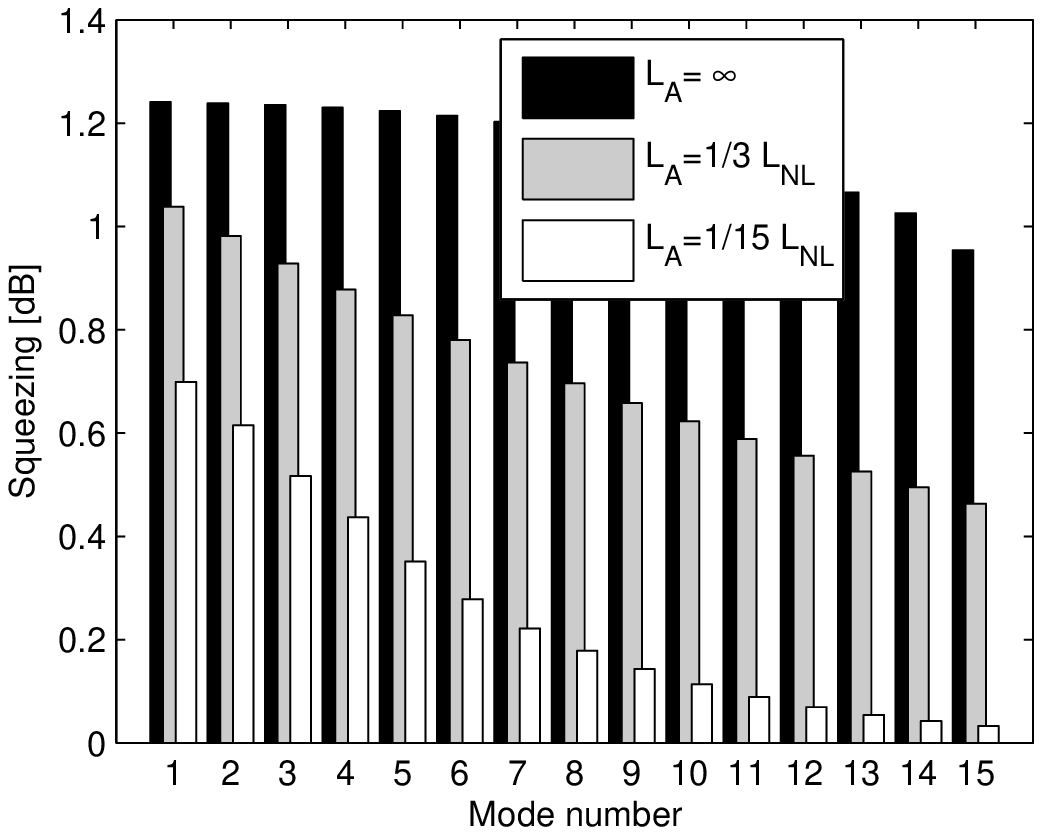}} 
\caption{(a) The most squeezed function modulus (solid line) and phase (dashed line). (b) Histogram of the noise reductions versus mode number.}
\end{figure}

We found the most squeezed mode function for the parameters used in the Sec. 3: $L_{OV}=1/10 L_{NL}$, $L_D=3 L_{NL}$, $L_A=1/5 L_{NL}$, $L=L_{NL}$. The mode function presented in Fig. \ref{fig:fmod} is close to Gaussian. In the Fig. \ref{fig:mbar} we also present the squeezing versus mode number. The modes are found as subsequent eigenvectors of the quadratic form corresponding to  \eqref{hom1}, which physically represent uncorrelated, squeezed quadratures of SPDC, as detailed in the Appendix.
In a typical lossless case on the output of long crystals  we obtain a number of modes exhibiting similar squeezing. Introduction of losses decreases squeezing most rapidly in higher order modes. 

Let us analyse a quadrature conjugate to  $\hat{x}$,  $\hat{x}_{\pi/2}=\frac{i}{\sqrt{2}}\int \dd \omega \ \left[ -\lo^*(\omega) \ao{}{\omega}+\lo(\omega) \axo{}{\omega} \right] $. In the lossless case $ \hat{x}_{\pi/2}$ is the most antisqueezed quadrature, and the product $\xx\xy=1/4$ saturates the uncertainty bound.   However in the presence of losses the  product $\xx\xy$ grows during the evolution. In the lossy case this quantity can be treated as a measure of the impurity of the state and we will track its evolution. We have checked that in the lossy case the most antisqueezed quadrature differs from the quadrature conjugate to the most squeezed one. In addition to examining the evolution of $\xx$ and $\xx\xy$ we will track the evolution of average number of photons in the SPDC $\langle \hat{n} \rangle = \int \dd  \omega \ \avg{\hat{a}^\dagger (\omega) \hat{a}(\omega)}$. 

\begin{figure}
 \centering
 \includegraphics[width=16 cm,bb=0 0 513 357]{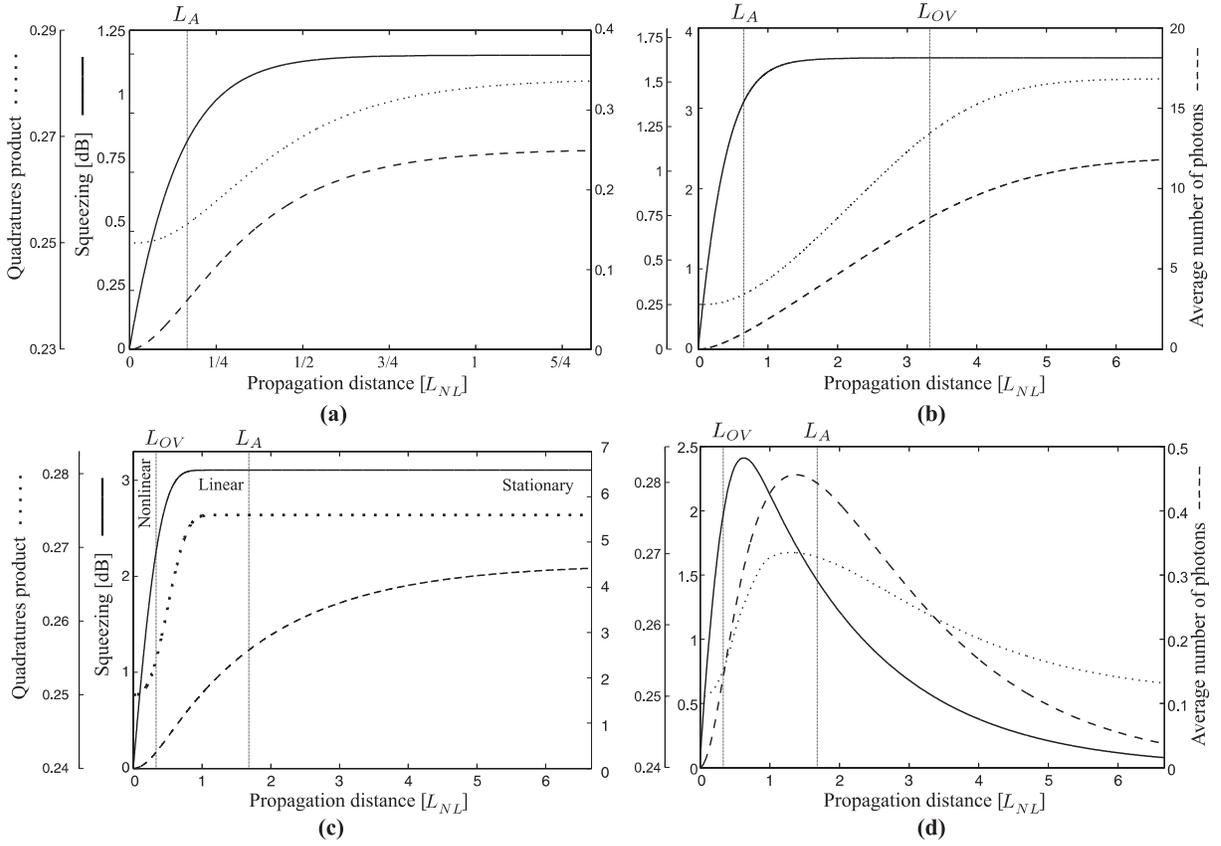}
 \caption{Plots presenting the squeezing, the average number of photons and the product of quadratures as the function of the propagation distance expressed in $L_{NL}$ units. In the plots we denoted the pump-SPDC overlap length $L_{OV}$ and the attenuation length $L_A$ which values determine the type of the evolution. (a) The stationary state in the regime of big losses $L_A<L_{NL}/2$ with no dispersion $L_{OV}\rightarrow\infty$ (b) Stabilization of squeezing with the simultaneous growth of average number of photons and the quadrature product $\xx\xy$ on the effective crystal length $L<L_{OV}$. (c) Typical realistic case where $L_A>L_{OV}$ leads to the stationary states of the correlation functions. Three stages of the evolution: nonlinear, linear and stationary. (d) Adiabatic attenuation of all tracked quantities from (c) with the pump dumping enabled.}
 \label{sqzev}
\end{figure}

Now we will be tracking the evolution of $\xx$, $\xx\xy$ and  $\avg{\hat{n}}$ for different values of the parameters. Firstly, let us take the case with neglectable dispersion $L_{OV},L_D\rightarrow \infty$ when the one mode solution \eqref{eq:analsol} can be applied. As it was discussed above, depending on the relation between $L_A$ and $L_{NL}/2$,  different types of evolution occur. In the Fig \ref{sqzev}(a) we present the big losses regime for $L_A=1/6 \ L_{NL}<1/2 \ L_{NL}$. For $L<L_A$ squeezing $\xx$, number of photons $\avg{\hat n}$ and the product of quadratures $\xx\xy$ grow. Afterwards losses stabilize all of them for $L>L_A$. Physically the attenuation balances a nonlinear amplification. Let us notice that regardless of the attenuation in the medium, one can always find squeezed quadrature in SPDC.  Since losses continuously add a vacuum noise to the SPDC, and the mode becomes more classical as $\xx\xy$ grows, it is suprising that we always find some nonclassical properties in the output SPDC light.

For $L_A> L_{NL}/2$ without any linear propagation, we encounter a stabilization of squeezing as in the previous case. Interestingly both $\avg{\hat n}$ and $\xx\xy$ grow. The most squeezed quadrature remains constant but reduced below the shot noise level but its conjugate quadrature grows unlimitedly. When we deal with a  more realistic situation where  $L_{OV}$ is finite, the growth continues until $L<L_{OV}$ as presented in the  Fig. \ref{sqzev}(b). For longer crystals we observe a stationary state where the other two quantities $\left \langle \hat{n} \right \rangle$ and $\xx\xy$ stabilize.

Typically the losses in the material are relatively small and we have $L_A>L_{OV}$. This situation is presented in the Fig. \ref{sqzev}(c). Here, the initial evolution of the SPDC light resembles the lossless case and it can be divided into two stages. During the first stage we observe the nonlinear growth of the average photon number and squeezing until $L<L_{OV}$. In the next stage squeezing stabilizes and the SPDC process generates many similarly squeezed modes. Thus the average number of photons increases linearly. The effects of losses can be seen in the third, stationary stage, when $L>L_A$. In this stage we reach the stabilization of all tracked quantities. In the two previously discussed cases, the stabilization emerged when the nonlinear interaction still played an important role. Here we deal with the attenuation of tails of the CF, described in details in Sec. 3 (Fig. \ref{fig:cf}), where the multimode CF  achieve the stationary state.

In all of the discussed cases we neglected the second order dispersion $L_D\rightarrow\infty$. We have checked that the finite dispersion length $L_D$ does not influence the noise reduction in the most squeezed mode or the total average number of photons. Moreover, for a real media we usually have $L_D \gg L_{OV}$ and the ratio between $L_{NL}, \ L_{OV}$ and $L_A$ plays dominant role.

In the end we present what happens if we take into account both the pump and the SPDC light attenuation. Besides of the pump losses, in the Fig. \ref{sqzev}(d) the other parameters are the same as in the Fig. \ref{sqzev}(c). Since the pump seeds the SPDC process its attenuation prevents reaching the balance between the losses and the nonlinear light generation. In the consequence the CF decrease adiabatically, together with the squeezing and the average number of photons.

\section{Summary}

In this paper we studied the process of the multimode, one dimensional Spontaneous Parametric Down--Conversion in the lossy media. We utilized relatively simple approach to the problem consisting in  analysis of the propagation of the first order Correlation Functions (CF) of the fluorescence fields. These functions provide a complete information about the Gaussian multimode state of light.

We wrote equations of  propagation of the CF and expressed them using four characteristic length scales. We showed that the evolution of the CF can be qualitatively predicted, and the character of the solutions is fully determined by the ratios of the four length scales. In particular, in case of negligible dispersion the multimode problem simplifies to the one mode evolution, for which we found analytical solutions. We further differentiated the regimes of high and low losses for which we predict either a stabilization or an exponential growth of average number of photons in the SPDC light. The presence of losses stabilizes the most squeezed quadrature. However in the case of the small losses the antisqueezed quadrature grows boundlessly. 
 
In the multimode case we found that losses make the simple picture of separable modes of light not longer valid.  However we are still able to single out uncorrelated squeezed quadratures.  We described a simple procedure to find corresponding orthogonal modes. 
We tracked the evolution of the average number of photons in the SPDC. In addition we calculated the highest available squeezing and a product of the squeezing and the antisqueezing in the corresponding mode which we use as a measure of the impurity of the state. Examination of the development of these three quantities allowed us to divide the process into three general stages: an exponential growth and a squeezing due to a nonlinear interaction, a linear temporal expansion due to a group velocity mismatch and a stabilization due to losses. Those stages are well separated in a typical case when group velocity mismatch between pump and SPDC plays significant role on shorter distances than the attenuation does. 

We wish to emphasize that losses during the development of the spontaneous fluorescence light do not entirely destroy its quantum properties and the state always remains squeezed below the shot noise level. Under certain conditions this happens although both the number of photons in SPDC and the antisqueezing grow.
 
 Despite the fact the whole process contains the simultaneous effects of a squeezing, mode mixing and attenuation, we were able to divide the evolution into simple stages and predict it qualitatively. We find the description using multimode correlation functions very fruitful and promising. This approach could be also applied to other quantum problems where we observe simultaneous spontaneous scattering and losses. Possibilities include a nondegenerate three dimensional OPA \cite{Chwedenczuk08}, Raman scattering in atomic vapors \cite{Raymer1DTheory81} and modelling of the collisions of BECs where spontaneously scattered atoms are observed \cite{Norrie05}.

\section{Acknowledgments}
We acknowledge insightful discussions with Konrad Banaszek and Czes{\l}aw Radzewicz,
as well as financial support from Polish Government scientific grant (2007-2009). 

\appendices

\section*{Appendix}
In the absence of losses the optimal mode minimizing \eqref{hom1} is the eigenmode of $\aaw$ with highest eigenvalue. Moreover we can find a set of uncorrelated modes squeezed independently. Unfortunately in the presence of losses neither is true. 
We have to minimise a general expression \eqref{hom1} by proper choice of the local oscillator mode $\lo(\w)$. This is easily accomplished using linear-algebraic means. Firstly let us notice that the variance $\left \langle \hat{x}^2 \right \rangle$ in \eqref{hom1} is real and can be reexpressed in the following symmetrized way:
 \begin{multline}
 \label{homappsen}
  \left \langle \hat{x}^2 \right \rangle=
  \frac{1}{2} +\iint  \dd \omega \dd \omega' \left[ \frac{1}{2}\left(\lo(\omega) \axaw \lo^*(\omega')+ \right. \right.  \\
  \left. \left. \lo^*(\omega) \axaw^* \lo(\omega') \right) + \re\left(\lo(\omega) \aaw \lo(\omega') \right) \right]
\end{multline}

As we apply a matrix approximation to the CF $\axaw, \ \aaw$ and vector approximation to $\lo(\w)$ the integration in \eqref{homappsen} becomes multiplication of a matrix by two vectors. The product is sensitive to the total phase of $\lo(\w)$ and therefore one has to represent it by a vector consisting of the real and imaginary part of $\lo$:
\begin{equation}
  \mathbf{\Phi}=
  \begin{bmatrix}
{\re(\lo)}\\
 {\im(\lo)}
\end{bmatrix}.
\end{equation}
Then we can rewrite \eqref{homappsen} in the following form:
\begin{equation}
 \left \langle \hat{x}^2 \right \rangle =1/2+\mathbf{\Phi}^T\mathbf{M}\mathbf{\Phi}
 \label{quadform3}
\end{equation}
where $\mathbf{M}$ is the matrix of the quadratic form, expressed by the matrix approximations of CF $\axaw$ and $\aaw$:
\begin{equation}
 \mathbf{M}=
 \begin{bmatrix}  
    \re\left(  \gc\right) +\re\left(  \fc\right), &
   - \im\left(  \fc\right)+\im\left(  \gc\right)\\
      -\im\left(  \fc\right) - \im\left(  \gc\right), &
  \re\left(  \gc\right) -\re\left(  \fc\right) \\
 \end{bmatrix}.
\end{equation}
Since $\im\left(\gc\right)$ is antisymmetric, $\mathbf{M}$ is real and symmetric and its eigenvectors $\mathbf{\Phi}_i$ are orthogonal. 
Therefore the correlation between two quadratures $\hat x_i$ and $\hat x_j$ corresponding to different eigenmodes $\mathbf{\Phi}_i$ and $\mathbf{\Phi}_j$ is zero:
\begin{equation}
\langle \hat{x}_i \hat{x}_j \rangle=\mathbf{\Phi}^T_i\mathbf{M}\mathbf{\Phi}_j=0, \quad i\neq j.
\end{equation}

\bibliographystyle{tMOP}
\bibliography{all}

\begin{thebibliography}{15}
\providecommand{\natexlab}[1]{#1}

\bibitem[1]{Furusawa98}
Furusawa, A.; Sorensen, J.L.; Braunstein, S.L.; Fuchs, C.A.; Kimble, H.J.;
  et~al. Unconditional Quantum Teleportation.  {\em Science}  {\bf 1998}, {\em
  282}, 706.

\bibitem[2]{RalphPRA00}
Ralph, T.C. Continuous variable quantum cryptography.  {\em Phys. Rev. A}  {\bf
  1999}, {\em 61}, 010303(R).

\bibitem[3]{VahlbruchPRL08}
Vahlbruch, H.; Mehmet, M.; Chelkowski, S.; Hage, B.; Franzen, A.; Lastzka, N.;
  Gossler, S.; Danzmann, K.; et~al. Observation of Squeezed Light with 10-dB
  Quantum-Noise Reduction.  {\em Phys. Rev. Lett.}  {\bf 2008}, {\em 100} (3),
  033602.

\bibitem[4]{KuzmichPRL00}
Kuzmich, A.; Mandel, L.; Bigelow, N.P. Generation of Spin Squeezing via
  Continuous Quantum Nondemolition Measurement.  {\em Phys. Rev. Lett.}  {\bf
  2000}, {\em 85}, 1594.

\bibitem[5]{SchoriPRA02}
Schori, C.; S\o{}rensen, J.L.; Polzik, E.S. Narrow-band frequency tunable light
  source of continuous quadrature entanglement.  {\em Phys. Rev. A}  {\bf
  2002}, {\em 66}, 033802.

\bibitem[6]{AndersonOL95}
Anderson, M.E.; Beck, M.; Raymer, M.G.; et~al. Quadrature squeezing with
  ultrashort pulses in nonlinear-optical waveguides.  {\em Opt. Lett.}  {\bf
  1995}, {\em 20} (6), 620--622.

\bibitem[7]{WasilewskiPRA06A}
Wasilewski, W.; Lvovsky, A.I.; Banaszek, K.; et~al. Pulsed squeezed light:
  Simultaneous squeezing of multiple modes.  {\em Phys. Rev. A}  {\bf 2006},
  {\em 73} (6), 063819.

\bibitem[8]{Keller97}
Keller, T.E.; Rubin, M.H. Theory of two-photon entanglement for spontaneous
  parametric down-conversion driven by a narrow pump pulse.  {\em Phys. Rev. A}
   {\bf 1997}, {\em 56}, 1534.

\bibitem[9]{BraunsteinBM}
Braunstein, S.L. Squeezing as an irreducible resource.  {\em Phys. Rev. A}
  {\bf 2005}, {\em 71}, 055801.

\bibitem[10]{OpatrnyPRA2002}
Opatrny, T.; Korolkova, N.; Leuchs, G. Mode structure and photon number
  correlations in squeezed quantum pulses.  {\em Phys. Rev. A}  {\bf 2002},
  {\em 66}, 53813.

\bibitem[11]{CarusoNJP08}
Caruso, F.; Eisert, J.; Giovannetti, V.; et~al. Multi-mode bosonic Gaussian
  channels.  {\em New Journal of Physics}  {\bf 2008}, {\em 10} (8), 083030.

\bibitem[12]{BoydNLO}
Boyd, R.W. Nonlinear Optics.  : , 1992.

\bibitem[13]{Chwedenczuk08}
Chwedenczuk, J.; Wasilewski, W. Intensity of parametric fluorescence pumped by
  ultrashort pulses.  {\em Phys. Rev. A}  {\bf 2008}, {\em 78} (063823).

\bibitem[14]{Raymer1DTheory81}
Raymer, M.G.; Mostowski, J. Stimulated {Raman} scattering: unified treatment of
  spontaneous initiation and spatial propagation.  {\em Phys. Rev. A}  {\bf
  1981}, {\em 24}, 1980.

\bibitem[15]{Norrie05}
A.~A.~Norrie, R.J.B.; Gardiner, C.W. Quantum Turbulence in Condensate
  Collisions: An Application of the Classical Field Method.  {\em Phys. Rev.
  Lett.}  {\bf 2005}, {\em 94} (040401).

\end{thebibliography}

\end{document}